\documentclass{article}%
\usepackage{amsfonts}
\usepackage{amsmath}
\usepackage{amssymb}
\usepackage{graphicx}%
\providecommand{\U}[1]{\protect\rule{.1in}{.1in}}
\newtheorem{theorem}{Theorem}
\newtheorem{acknowledgement}[theorem]{Acknowledgement}

\begin{document}

\title{Generalized Poincare algebras and Lovelock-Cartan gravity theory}
\author{P. K. Concha, D. M. Pe\~{n}afiel, E. K. Rodr\'{\i}guez, P. Salgado\\Departamento de F\'{\i}sica, Universidad de Concepci\'{o}n \\Casilla 160-C, Concepci\'{o}n, Chile}
\maketitle

\begin{abstract}
We show that the Lagrangian for Lovelock-Cartan gravity theory can be
re-formulated as an action which leads to General Relativity in a certain
limit. \ In odd dimensions the Lagrangian leads to a Chern-Simons theory
invariant under the generalized Poincar\'{e} algebra $\mathfrak{B}_{2n+1},$
while\ in even dimensions the Lagrangian leads to a Born-Infeld theory
invariant under a subalgebra of the $\mathfrak{B}_{2n+1}$ algebra. \ It is
also shown that torsion may occur explicitly in the Lagrangian leading to new
torsional Lagrangians, which are related to the Chern-Pontryagin character for
the $B_{2n+1}$ group.

\end{abstract}

\section{\textbf{Introduction}}

The most general metric theory of gravity satisfying the criteria of general
covariance and yielding to second-order field equations is a polynomial of
degree $\left[  d/2\right]  $ in the curvature known as the Lanczos-Lovelock
gravity theory (LL) \cite{lanc, lovel}. \ The LL action can be written as the
most general $d$-form invariant under local Lorentz transformations,
constructed with the spin connection, the vielbein and their exterior
derivatives, without the Hodge dual \cite{zum, teit},%
\begin{equation}
S=\int\sum_{p=0}^{\left[  d/2\right]  }\tilde{\alpha}_{p}\varepsilon
_{a_{1}a_{2}\cdots a_{d}}R^{a_{1}a_{2}}\cdots R^{a_{2p-1}a_{2p}}e^{a_{2p+1}%
}\cdots e^{a_{d}},
\end{equation}
where $R^{ab}=d\omega^{ab}+\omega_{\text{ }c}^{a}\omega^{cb}$ is the Lorentz
curvature, $e^{a}$ corresponds to the one-form vielbein and the coefficientes
$\tilde{\alpha}_{p}$, $p=0,1,\dots,\left[  d/2\right]  ,$ are arbitrary
constants and they are not fixed from first principles.

It is an accepted fact that requiring the LL theory to have the maximum
possible number of degrees of freedom, fixes the parameters $\tilde{\alpha
}_{p}$'s in terms of the gravitational and the cosmological constants
\cite{tron}. \ As a consequence, the action in odd dimensions can be
formulated as a Chern-Simons ($ChS$) theory of the $AdS$ group, while in even
dimensions the action has a Born-Infeld ($BI$) form invariant only under local
Lorentz rotations in the same way as the Einstein-Hilbert action \cite{tron,
ban1, ban2, zanel}.

Although the Einstein-Hilbert term is contained in the LL action, the $ChS$
gravity for the $AdS$ group and the $BI$ gravity for the Lorentz group are
dynamically very different from standard General Relativity.

In Ref. \cite{salg1} it was shown that the standard, odd-dimensional General
Relativity can be obtained from a Chern-Simons gravity theory for a certain
$\mathfrak{B}_{m}$ Lie algebra, which will be called generalized Poincar\'{e}
algebra%
\footnote{Alternatively known as the Maxwell algebra type.}
(where the particular case $\mathfrak{B}_{4}$ corresponds to the so-called
Maxwell algebra\ \cite{salg1c}). The generalized Poincar\'{e} algebras can be
obtained \ by a resonant reduced $S$-expansion of the $AdS$ Lie algebra using
$S_{E}^{(N)}=\left\{  \lambda_{\alpha}\right\}  _{\alpha=0}^{N+1}$ as
semigroup \cite{salg1}.\ \ 

The $S$-expansion method has been introduced in Ref. \cite{salg2} (see also
\cite{hat}, \cite{azc1}, \cite{azc2} ) and consists in a powerfull tool in
order to obtain new Lie algebras from original ones. \ The method is based on
combining the structure constants of a Lie algebra $\mathfrak{g}$ with the
inner multiplication law of a semigroup $S$. \ The new Lie algebra
$\mathfrak{G}=S\times\mathfrak{g}$ is called the $S$-expanded algebra.
\ Interestingly, when a decomposition of the semigroup $S=%
{\displaystyle\bigcup\nolimits_{p\in I}}
S_{p}$ $\left(  \text{where }I\text{ is a set of indices}\right)  $ satisfies
the same structure that the subspaces $V_{p}$ of the original algebra
$\mathfrak{g}=%
{\displaystyle\bigoplus\nolimits_{p\in I}}
V_{p}$, we say that $\mathfrak{G}_{R}=%
{\displaystyle\bigoplus\nolimits_{p\in I}}
S_{p}\times V_{p}$ is a resonant subalgebra of $\mathfrak{G=}S\times g$. \ In
particular, when the semigroup has a zero element $0_{S}$, the reduced algebra
is obtained imposing $0_{S}\times\mathfrak{g}=0$.

Subsequently, in Ref. \cite{salg1a} it was found that\ standard
even-dimensional General Relativity emerges as a limit of a Born-Infeld theory
invariant under a certain subalgebra $\mathfrak{L}^{\mathfrak{B}_{m}}$ of the
$\mathfrak{B}_{m}$ Lie algebra. \ These odd- and even-dimensional theories are
described by the so called Einstein-Chern-Simons ($EChS$) and the
Einstein-Born-Infeld ($EBI$) actions, respectively.

Very recently it was found in Ref. \cite{CPRS2} that\ standard odd and
even-dimensional General Relativity emerges\textbf{ }as a weak coupling
constant limit of a $(2p+1)$-dimensional Chern-Simons Lagrangian and of a
$2p$-dimensional Born-Infeld Lagrangian invariant under $\mathfrak{B}_{2m+1}$
and $\mathfrak{L}^{\mathfrak{B}_{2m}}$, respectively, if and only if $m\geq p$.

It is the purpose of this paper to show that: $(i)$\ it is possible to
reformulate the Lagrangian for Lovelock-Cartan gravity theory, which we call
"Lagrangian of $Einstein$-$Lovelock$-$Cartan$ ($ELC$)", such that, in odd
dimensions leads to the Einstein-Chern-Simons Lagrangian, and in even
dimensions leads to the Einstein-Born-Infeld Lagrangian; $(ii)$ the torsion
may occur explicitly in the Lagrangian and that, following a procedure
analogous to that of Ref. \ \cite{tron}, it is possible to find new torsional
Lagrangians, which are related to the Chern-Pontryagin character for the
$\mathfrak{B}_{2n+1}$ group.

This paper is organized as follows: In Section~$2$ we briefly review some
aspects of the construction of the so called generalized Poincar\'{e} algebras
and how it is possible to obtain General Relativity from the Chern-Simons and
Born-Infeld formalism using these algebras.

In Section $3$ the $ELC$-Lagrangian is constructed. It is shown that this
Lagrangian leads in odd dimensions to the $EChS$ Lagrangian and in even
dimensions leads to the $EBI$ Lagrangian.

In Section $4$ the $ELC$-Lagrangian is generalized adding torsion explicitly
following a procedure analogous to that of Ref. \ \cite{tron}. It is shown
that in $4p$ dimensions, the only $4p$-forms $\mathfrak{B}_{2n+1}$-invariant,
constructed from $e^{\left(  a,2k+1\right)  }$, $R^{\left(  ab,2k\right)  }$
and $T^{\left(  a,2k+1\right)  }$ $\left(  k=0,\cdots,n-1\right)  $, are
Pontryagin type invariants $P_{\left(  4p\right)  }$.

In Section $5$ we show that using the dual formulation of the $S$-expansion
introduced in Ref. \cite{salg3}, it is possible to relate the Euler type
invariant and the Pontryagin type invariant in $d=3$ dimensions. Section $6$
concludes the work with a comment and possible developments. \ 

\section{\textbf{General Relativity and the generalized Poincar\'{e} algebras
}$\mathfrak{B}_{2n+1}$}

In order to describe how the action for General Relativity can be obtained
from the gravity actions invariant under generalized Poincar\'{e} algebras,
let us review here the results obtained in Refs. \cite{salg1, salg1a, CPRS2}.
\ Following the definitions of Ref. \cite{salg2} let us consider
the\ $S$-expansion of the Anti-de Sitter $\left(  AdS\right)  $ Lie algebra
using as a semigroup $S_{E}^{\left(  2n-1\right)  }=\left\{  \lambda
_{0},\cdots,\lambda_{2n}\right\}  $ endowed with the multiplication law
$\lambda_{\alpha}\lambda_{\beta}=\lambda_{\alpha+\beta}$ \ when $\alpha
+\beta\leq2n$; \ $\lambda_{\alpha}\lambda_{\beta}=$ $\lambda_{2n}$\ when
\ $\alpha+\beta>2n$. \ The $\tilde{J}_{ab}$, $\tilde{P}_{a}$ generators of the
$AdS$ algebra satisfy the following commutation relations%
\begin{align}
\left[  \tilde{J}_{ab},\tilde{J}_{cd}\right]   &  =\eta_{bc}\tilde{J}%
_{ad}-\eta_{ac}\tilde{J}_{bd}-\eta_{bd}\tilde{J}_{ac}+\eta_{ad}\tilde{J}%
_{bc},\\
\left[  \tilde{J}_{ab},\tilde{P}_{c}\right]   &  =\eta_{bc}\tilde{P}_{a}%
-\eta_{ac}\tilde{P}_{b},\\
\left[  \tilde{P}_{a},\tilde{P}_{b}\right]   &  =\tilde{J}_{ab},
\end{align}
where $a,b=0,\dots,2n$ and $\eta_{ab}$ corresponds to the Minkowski metric.
Let us consider the following subset decomposition $S_{E}^{\left(
2n-1\right)  }=S_{0}\cup S_{1}$, with%
\begin{align}
S_{0}  &  =\left\{  \lambda_{2m},\text{ with }m=0,\ldots,n-1\right\}
\cup\left\{  \lambda_{2n}\right\}  ,\\
S_{1}  &  =\left\{  \lambda_{2m+1},\text{ with }m=0,\ldots,n-1\right\}
\cup\left\{  \lambda_{2n}\right\}  ,
\end{align}
where $\lambda_{2n}$ corresponds to the zero element of the semigroup $\left(
0_{S}=\lambda_{2n}\right)  $. \ After extracting a resonant subalgebra and
performing its $0_{S}$ $\left(  =\lambda_{2n}\right)  $-reduction, one finds
the generalized Poincar\'{e} algebra $\mathfrak{B}_{2n+1}$,
\begin{align}
\left[  P_{a},P_{b}\right]   &  =Z_{ab}^{\left(  1\right)  },\text{
\ \ \ \ }\left[  J_{ab},P_{c}\right]  =\eta_{bc}P_{a}-\eta_{ac}P_{b},\\
\left[  J_{ab,}J_{cd}\right]   &  =\eta_{cb}J_{ad}-\eta_{ca}J_{bd}+\eta
_{db}J_{ca}-\eta_{da}J_{cb},\label{sub1}\\
\left[  J_{ab},Z_{c}^{\left(  i\right)  }\right]   &  =\eta_{bc}Z_{a}^{\left(
i\right)  }-\eta_{ac}Z_{b}^{\left(  i\right)  },\\
\left[  Z_{ab}^{\left(  i\right)  },P_{c}\right]   &  =\eta_{bc}Z_{a}^{\left(
i\right)  }-\eta_{ac}Z_{b}^{\left(  i\right)  },\\
\left[  Z_{ab}^{\left(  i\right)  },Z_{c}^{\left(  j\right)  }\right]   &
=\eta_{bc}Z_{a}^{\left(  i+j\right)  }-\eta_{ac}Z_{b}^{\left(  i+j\right)
},\\
\left[  J_{ab,}Z_{cd}^{\left(  i\right)  }\right]   &  =\eta_{cb}%
Z_{ad}^{\left(  i\right)  }-\eta_{ca}Z_{bd}^{\left(  i\right)  }+\eta
_{db}Z_{ca}^{\left(  i\right)  }-\eta_{da}Z_{cb}^{\left(  i\right)
},\label{sub2}\\
\left[  Z_{ab,}^{\left(  i\right)  }Z_{cd}^{\left(  j\right)  }\right]   &
=\eta_{cb}Z_{ad}^{\left(  i+j\right)  }-\eta_{ca}Z_{bd}^{\left(  i+j\right)
}+\eta_{db}Z_{ca}^{\left(  i+j\right)  }-\eta_{da}Z_{cb}^{\left(  i+j\right)
},\label{sub3}\\
\left[  P_{a},Z_{c}^{\left(  i\right)  }\right]   &  =Z_{ab}^{\left(
i+1\right)  },\text{ \ \ \ \ }\left[  Z_{a}^{\left(  i\right)  }%
,Z_{c}^{\left(  j\right)  }\right]  =Z_{ab}^{\left(  i+j+1\right)  },
\end{align}
where $i,j=1,\cdots,n-1$. \ Let us note that the generators of the
$\mathfrak{B}_{2n+1}$ algebra are related to the original ones through%
\begin{align}
J_{ab}  &  =J_{\left(  ab,0\right)  }=\lambda_{0}\otimes\tilde{J}%
_{ab},\label{gen1}\\
P_{a}  &  =P_{\left(  a,1\right)  }=\lambda_{1}\otimes\tilde{P}_{a},\\
Z_{ab}^{\left(  i\right)  }  &  =J_{\left(  ab,2i\right)  }=\lambda
_{2i}\otimes\tilde{J}_{ab},\\
Z_{a}^{\left(  i\right)  }  &  =P_{\left(  a,2i+1\right)  }=\lambda
_{2i+1}\otimes\tilde{P}_{a}, \label{gen4}%
\end{align}
then if $i>n-1$ we have $Z_{ab}^{\left(  i\right)  }=Z_{a}^{\left(  i\right)
}=0$. \ The generalized Poincar\'{e} algebra $\mathfrak{B}_{2n+1}$ is also
known as the Maxwell algebra type which was introduced in Ref. \cite{CPRS2}.
\ We note that setting $Z_{ab}^{\left(  i+1\right)  }$ and $Z_{a}^{\left(
i\right)  }$ equal to zero, we obtain the $\mathfrak{B}_{4}$ algebra which
coincides with the Maxwell algebra $\mathcal{M}$ \cite{salg1c}. \ In fact,
every generalized Poincar\'{e} algebra $\mathfrak{B}_{l}$ can be obtained from
$\mathfrak{B}_{2n+1}$ setting some generators equal to zero. \ Besides, one
can see that the commutators $\left(  \ref{sub1}\right)  $, $\left(
\ref{sub2}\right)  $ and $\left(  \ref{sub3}\right)  $ form a Lorentz type
subalgebra of the $\mathfrak{B}_{2n+1}$ algebra. \ This subalgebra denoted as
$\mathfrak{L}^{\mathfrak{B}_{2n+1}}$ can be obtained as an $S$-expansion of
the Lorentz algebra $\mathfrak{L}$ using $S_{0}^{\left(  2n-1\right)
}=\left\{  \lambda_{0},\lambda_{2},\lambda_{4},\dots,\lambda_{2n}\right\}  $
as the relevant semigroup \cite{salg1a}.

The generalized Poincar\'{e} algebras are particularly interesting in the
context of gravity since it was shown in \cite{salg1} that standard
odd-dimensional General Relativity may emerge as the weak coupling constant
limit $\left(  l\rightarrow0\right)  $ of a $\left(  2n+1\right)
$-dimensional Chern-Simons Lagrangian invariant under the $\mathfrak{B}%
_{2n+1}$ algebra,%
\begin{align}
L_{CS\text{ \ }(2n+1)}^{\mathfrak{B}_{2n+1}}  &  =\sum_{k=1}^{n}l^{2k-2}%
c_{k}\alpha_{j}\delta_{i_{1}+\cdots+i_{n+1}}^{j}\delta_{p_{1}+q_{1}}^{i_{k+1}%
}\cdots\delta_{p_{n-k}+q_{n-k}}^{i_{n}}\varepsilon_{a_{1}\cdots a_{2n+1}%
}\nonumber\\
&  \times R^{\left(  a_{1}a_{2},i_{1}\right)  }\cdots R^{\left(
a_{2k-1}a_{2k},i_{k}\right)  }e^{\left(  a_{2k+1},p_{1}\right)  }e^{\left(
a_{2k+2},q_{1}\right)  }\cdots\nonumber\\
&  \cdots e^{\left(  a_{2n-1},p_{n-k}\right)  }e^{\left(  a_{2n}%
,q_{n-k}\right)  }e^{\left(  a_{2n+1},i_{n+1}\right)  }, \label{ECS}%
\end{align}
where%
\begin{align*}
c_{k}  &  =\frac{1}{2(n-k)+1}\left(
\begin{array}
[c]{c}%
n\\
k
\end{array}
\right) \\
R^{\left(  ab,2i\right)  }  &  =d\omega^{\left(  ab,2i\right)  }+\eta
_{cd}\omega^{\left(  ac,2j\right)  }\omega^{\left(  db,2k\right)  }%
\delta_{j+k}^{i},
\end{align*}
and $\alpha_{j}$ are arbitrary constants which appear as a consequence of the
$S$-expansion process.\ \ \ Let us note that the $S$-expanded fields are
related to the $AdS$ fields $\left\{  \tilde{e}^{a},\tilde{\omega}%
^{ab}\right\}  $ as follows,%
\begin{align*}
e^{\left(  a2j+1\right)  }  &  =\lambda_{2j+1}\otimes\tilde{e}^{a},\\
\omega^{\left(  ab,2j\right)  }  &  =\lambda_{2j}\otimes\tilde{\omega}^{ab},
\end{align*}
where $j=0,1,\dots,n-1$. $\ $In a similar way, the $S$-expanded Lorentz
curvature $R^{\left(  ab,2i\right)  }$ is related to the Lorentz curvature
$\tilde{R}^{ab}=d\tilde{\omega}^{ab}+\tilde{\omega}_{\text{ }c}^{a}%
\tilde{\omega}^{cb}$ as $R^{\left(  ab,2i\right)  }=\lambda_{2i}\tilde{R}%
^{ab}$.

Similarly, it was shown in \cite{salg1a} that standard even-dimensional
General Relativity emerges as the weak coupling constant limit $\left(
l\rightarrow0\right)  $ of a $\left(  2n\right)  $-dimensional Born-Infeld
type Lagrangian invariant under a subalgebra%
\footnote{The Lorentz type algebra $\mathfrak{L}^{{\mathfrak{B}}_{2n}%
}$ is identical to $\mathfrak{L}^{{\mathfrak{B}}_{2n+1}}$.}
$\mathfrak{L}^{\mathfrak{B}_{2n}}$ of the $\mathfrak{B}_{2n+1}$ algebra,%
\begin{align}
L_{BI\text{ \ }(2n)}^{\mathfrak{L}^{\mathfrak{B}_{2n}}}  &  =\sum_{k=1}%
^{n}l^{2k-2}\frac{1}{2n}\binom{n}{k}\alpha_{j}\delta_{i_{1}+\cdots+i_{n}}%
^{j}\delta_{p_{1}+q_{1}}^{i_{k+1}}\cdots\delta_{p_{n-k}+q_{n-k}}^{i_{n}%
}\nonumber\\
&  \varepsilon_{a_{1}\cdots a_{2n}}R^{\left(  a_{1}a_{2},i_{1}\right)  }\cdots
R^{\left(  a_{2k-1}a_{2k},i_{k}\right)  }e^{\left(  a_{2k+1},p_{1}\right)
}\nonumber\\
&  e^{\left(  a_{2k+2},q_{1}\right)  }\cdots e^{\left(  a_{2n-1}%
,p_{n-k}\right)  }e^{\left(  a_{2n},q_{n-k}\right)  }. \label{EBI}%
\end{align}

These results have recently been generalized in Ref. \cite{CPRS2} in which the
autors have shown that $L_{CS\text{ \ }(2n+1)}^{\mathfrak{B}_{2m+1}}$ and
$L_{BI\text{ \ }(2n)}^{\mathfrak{L}^{\mathfrak{B}_{2m}}}$ lead to the
Einstein-Hilbert Lagrangian in a weak coupling constant limit, if and only if
$m\geq n.$

\section{\textbf{The Einstein-Lovelock-Cartan Lagrangian}}

\qquad We have seen that the $S$-expansion procedure allows the construction
of Chern-Simons gravities in odd dimensions invariant under the $\mathfrak{B}%
_{2n+1}$ algebra and Born-Infeld type gravities in even dimensions invariant
under the $\mathfrak{L}^{\mathfrak{B}_{2n+1}}$ algebra, leading to General
Relativity in a certain limit. These gravities are called the
Einstein-Chern-Simons theories \cite{salg1} and the Einstein-Born-Infeld
theories \cite{salg1a}, respectively. These findings show that it could be
possible to reformulate the Lagrangian for Lovelock-Cartan gravity theory such
that, in a certain limit, it leads to the General Relativity theory.

In this section we show that it is possible to write a Lovelock-Cartan
Lagrangian leading to the $EChS$ Lagrangian in $d=2n-1$ invariant under the
$\mathfrak{B}_{2n-1}$ algebra, and to the $EBI$ Lagrangian in $d=2n$
\ invariant under the $\mathcal{L}^{\mathfrak{B}_{2n}}$ algebra. \ For this
purpose we shall use the useful properties of the $S$-expansion procedure
using $S_{E}^{\left(  d-2\right)  }$as the relevant semigroup.

The expanded action is given by%

\begin{equation}
S_{\mathcal{ELC}}=%
{\displaystyle\int}
{\displaystyle\sum\limits_{p=0}^{\left[  d/2\right]  }}
\mu_{i}\alpha_{p}L_{\mathcal{ELC}}^{\left(  p,i\right)  }\label{lovexp}%
\end{equation}
where $\alpha_{p}$ and $\mu_{i}$, with $i=0,...,d-2$, are arbitrary constants
and $L_{\mathcal{ELC}}^{\left(  p,i\right)  }$ is given by%
\begin{equation}
L_{\mathcal{ELC}}^{\left(  p,i\right)  }=l^{d-2}\delta_{i_{1}+\cdots+i_{d-p}%
}^{i}\varepsilon_{a_{1}a_{2}\cdots a_{d}}R^{\left(  a_{1}a_{2},i_{1}\right)
}\cdots R^{\left(  a_{2p-1}a_{2p},i_{p}\right)  }e^{\left(  a_{2p+1}%
,i_{p+1}\right)  }\cdots e^{\left(  a_{d},i_{d-p}\right)  },\label{EL}%
\end{equation}
with%
\begin{equation}
R^{\left(  ab,2i\right)  }=d\omega^{\left(  ab,2i\right)  }+\eta_{cd}%
\omega^{\left(  ac,2j\right)  }\omega^{\left(  db,2k\right)  }\delta_{j+k}%
^{i}.
\end{equation}
The expanded fields $\left\{  e^{\left(  a,2i+1\right)  },\omega^{\left(
ab,2i\right)  }\right\}  $ are related to the $AdS$ fields $\left\{  \tilde
{e}^{a},\tilde{\omega}^{ab}\right\}  $ as follows
\begin{align}
\omega^{\left(  ab,2i\right)  } &  =\lambda_{2i}\otimes\tilde{\omega}^{ab},\\
e^{\left(  a,2i+1\right)  } &  =\lambda_{2i+1}\otimes\tilde{e}^{a},
\end{align}
where $\lambda_{\alpha}\in$ $S_{E}^{\left(  d-2\right)  },$ which is a
semigroup that obey the following multiplication law (see Ref. \cite{salg2}),%
\begin{equation}
\lambda_{\alpha}\lambda_{\beta}=\left\{
\begin{array}
[c]{c}%
\lambda_{\alpha+\beta},\text{ \ \ when }\alpha+\beta\leq d-1,\text{
\ \ \ \ \ \ \ \ \ \ \ \ \ }\\
\lambda_{d-1},\text{ \ \ when \ }\alpha+\beta
>d-1.\text{\ \ \ \ \ \ \ \ \ \ \ \ \ }%
\end{array}
\right.
\end{equation}
Following the same procedure of Ref. \cite{tron}, we consider the variation of
the action with respect to $e^{\left(  a,i\right)  }$and $\omega^{\left(
ab,i\right)  }$. \ The variation of the action (\ref{lovexp}) leads to the
following equations:
\begin{align}
\varepsilon_{a}^{\left(  i\right)  } &  =%
{\displaystyle\sum\limits_{p=0}^{\left[  \left(  d-1\right)  /2\right]  }}
\mu_{i}\alpha_{p}\left(  d-2p\right)  \varepsilon_{a}^{\left(  p,i\right)
}=0\text{ },\\
\varepsilon_{ab}^{\left(  i\right)  } &  =%
{\displaystyle\sum\limits_{p=1}^{\left[  \left(  d-1\right)  /2\right]  }}
\mu_{i}\alpha_{p}p\left(  d-2p\right)  \varepsilon_{ab}^{\left(  p,i\right)
}=0,\label{ec1a}%
\end{align}
where%
\begin{align}
\varepsilon_{a}^{\left(  p,i\right)  }\colon &  =l^{d-2}\delta_{i_{1}%
+\cdots+i_{d-p-1}}^{i}\varepsilon_{ab_{1}\cdots b_{d-1}}R^{\left(  b_{1}%
b_{2},i_{1}\right)  }\cdots R^{\left(  b_{2p-1}b_{2p},i_{p}\right)
}\nonumber\\
&  \times e^{\left(  b_{2p+1},i_{p+1}\right)  }\cdots e^{\left(
b_{d-1},i_{d-p-1}\right)  },\label{ec1d}%
\end{align}%
\begin{align}
\varepsilon_{ab}^{\left(  p,i\right)  }\colon &  =l^{d-2}\delta_{i_{1}%
+\cdots+i_{d-p-1}}^{i}\varepsilon_{aba_{3}\cdots a_{d}}R^{\left(  a_{3}%
a_{4},i_{1}\right)  }\cdots R^{\left(  a_{2p-1}a_{2p},i_{p-1}\right)
}\nonumber\\
&  T^{\left(  a_{2p+1},i_{p}\right)  }e^{\left(  a_{2p+2},i_{p+1}\right)
}\cdots e^{\left(  a_{d},i_{d-p-1}\right)  },\label{ec1b}%
\end{align}
and where $T^{\left(  a,i\right)  }=de^{\left(  a,i\right)  }+\eta_{dc}%
\omega^{\left(  ad,j\right)  }e^{\left(  c,k\right)  }\delta_{j+k}^{i}$ is the
expanded $2$-form torsion. \ Using the covariant exterior derivative
$D=d+\left[  A,\cdot\right]  $ \ ( where $A$ corresponds to the one-form gauge
connection $\mathfrak{B}_{2n-1}$-valued) and the Bianchi identity for the
expanded $2$-form curvature $DR^{\left(  ab,i_{j}\right)  }=0$, we have%
\begin{align}
D\varepsilon_{a}^{\left(  p,i\right)  } &  =l^{d-2}\left(  d-1-2p\right)
\delta_{i_{1}+\cdots+i_{d-p-1}}^{i}\varepsilon_{ab_{1}\cdots b_{d-1}%
}R^{\left(  b_{1}b_{2},i_{1}\right)  }\cdots R^{\left(  b_{2p-1}b_{2p}%
,i_{p}\right)  }\nonumber\\
&  T^{\left(  b_{2p+1},i_{p+1}\right)  }e^{\left(  b_{2p+2},i_{p}\right)
}\cdots e^{\left(  a_{d-1},i_{d-p-1}\right)  }.\label{ec1}%
\end{align}
Since%
\begin{align}
e^{\left(  b,j\right)  }\varepsilon_{ba}^{\text{ \ \ \ }\left(  p,k\right)
}\delta_{j+k}^{i} &  =l^{d-2}\delta_{i_{1}+\cdots+i_{d-p-1}}^{i}%
\varepsilon_{aa_{1}\cdots a_{d-2}}R^{\left(  a_{1}a_{2},i_{1}\right)  }\cdots
R^{\left(  a_{2p-3}a_{2p-2},i_{p-1}\right)  }\nonumber\\
&  T^{\left(  a_{2p-1},i_{p}\right)  }e^{\left(  a_{2p},i_{p+1}\right)
}\cdots e^{\left(  a_{d-2},i_{d-p-1}\right)  },
\end{align}
one finds%
\begin{align}
e^{\left(  b,j\right)  }\varepsilon_{ba}^{\text{ \ \ \ }\left(  p+1,k\right)
}\delta_{j+k}^{i} &  =l^{d-2}\delta_{i_{1}+\cdots+i_{d-p}}^{i}\varepsilon
_{aa_{1}\cdots a_{d-1}}R^{\left(  a_{1}a_{2},i_{1}\right)  }\cdots R^{\left(
a_{2p-1}a_{2p},i_{p-1}\right)  }\nonumber\\
&  T^{\left(  a_{2p+1},i_{p}\right)  }e^{\left(  a_{2p+2},i_{p+1}\right)
}\cdots e^{\left(  a_{d-1},i_{d-p}\right)  }.\label{ec2}%
\end{align}
From $\left(  \ref{ec1}\right)  $ and $\left(  \ref{ec2}\right)  $ we have%
\[
D\varepsilon_{a}^{\left(  p,i\right)  }=\left(  d-1-2p\right)  e^{\left(
b,j\right)  }\varepsilon_{ba}^{\text{ \ \ \ }\left(  p+1,k\right)  }%
\delta_{j+k}^{i}%
\]
for $0\leq p\leq\left[  \left(  d-1\right)  /2\right]  $. \ This means that%
\begin{equation}
D\varepsilon_{a}^{\text{ \ }\left(  i\right)  }=%
{\displaystyle\sum\limits_{p=0}^{\left[  \left(  d-1\right)  /2\right]  }}
\mu_{i}\alpha_{p}\left(  d-2p\right)  \left(  d-1-2p\right)  e^{\left(
b,j\right)  }\varepsilon_{ba}^{\text{ \ \ \ }\left(  p+1,k\right)  }%
\delta_{j+k}^{i}.
\end{equation}
lf $p^{\prime}=p+1$ we find%
\begin{equation}
D\varepsilon_{a}^{\text{ \ }\left(  i\right)  }=%
{\displaystyle\sum\limits_{p^{\prime}=1}^{\left[  \left(  d+1\right)
/2\right]  }}
\mu_{i}\alpha_{p^{\prime}-1}\left(  d-2p^{\prime}+2\right)  \left(
d-2p^{\prime}+1\right)  e^{\left(  b,j\right)  }\varepsilon_{ba}^{\text{
\ \ \ }\left(  p^{\prime},k\right)  }\delta_{j+k}^{i},
\end{equation}
which can be rewritten as%
\begin{equation}
D\varepsilon_{a}^{\text{ \ }\left(  i\right)  }=%
{\displaystyle\sum\limits_{p=1}^{\left[  \left(  d+1\right)  /2\right]  }}
\mu_{i}\alpha_{p-1}\left(  d-2p+2\right)  \left(  d-2p+1\right)  e^{\left(
b,j\right)  }\varepsilon_{ba}^{\text{ \ \ \ }\left(  p,k\right)  }\delta
_{j+k}^{i},\label{ec3}%
\end{equation}
which by consistency with $\varepsilon_{a}^{\left(  i\right)  }=0$ must also
vanish. \ Taking the product of $\varepsilon_{ba}^{\text{ \ \ \ }\left(
k\right)  }$ with $e^{\left(  b,j\right)  }$ we find%
\begin{equation}
e^{\left(  b,j\right)  }\varepsilon_{ba}^{\text{ \ \ \ }\left(  k\right)
}\delta_{j+k}^{i}=%
{\displaystyle\sum\limits_{p=1}^{\left[  \left(  d-1\right)  /2\right]  }}
\mu_{i}\alpha_{p}p\left(  d-2p\right)  e^{\left(  b,j\right)  }\varepsilon
_{ba}^{\text{ \ \ \ }\left(  p,k\right)  }\delta_{j+k}^{i}\label{ec4}%
\end{equation}
which vanishes by consistency with $\varepsilon_{ab}^{\left(  i\right)  }=0$.

In general there are different ways of choosing the coefficients $\alpha_{p}$
which in general correspond to different theories with different numbers of
degrees of freedom. It is possible to choose the $\alpha_{p}$ such that
$\varepsilon_{a}^{\text{ \ }\left(  i\right)  }$ and $\varepsilon_{ab}^{\text{
\ }\left(  i\right)  }$ are independent. This last condition corresponds to
the maximum number of independent components.

\subsection{\textbf{Chern-Simons gravity invariant under }$\mathfrak{B}%
_{2n-1}$}

\qquad Following the same procedure of Ref. \cite{tron} one can see that in
the odd-dimensional case Eqs. $\left(  \ref{ec3}\right)  $, $\left(
\ref{ec4}\right)  $, lead to the coefficients given by%

\begin{equation}
\alpha_{p}=\alpha_{0}\frac{\left(  2n-1\right)  \left(  2\gamma\right)  ^{p}%
}{\left(  2n-2p-1\right)  }\binom{n-1}{p},
\end{equation}
where $\alpha_{0}$ and $\gamma$ are related to the gravitational and the
cosmological constants,
\begin{equation}
\alpha_{0}=\frac{\kappa}{\left(  l^{d-1}d\right)  };\text{ \ \ \ }%
\gamma=-sgn\left(  \Lambda\right)  \frac{l^{2}}{2}.
\end{equation}
\ For any dimension $d,$ $l$ is a length parameter related to the cosmological
constant by
\begin{equation}
\Lambda=\pm\frac{\left(  d-1\right)  \left(  d-2\right)  }{2l^{2}},
\end{equation}
and the gravitational constant $G$ is related to $\kappa$ through%
\begin{equation}
\kappa^{-1}=2\left(  d-2\right)  !\Omega_{d-2}G.
\end{equation}
With these coefficients the Lagrangian (\ref{lovexp}) may be written as the
Chern-Simons form%
\begin{align}
L_{CS\text{ \ }\left(  2n-1\right)  }^{\mathfrak{B}_{2n-1}}  &  =%
{\displaystyle\sum\limits_{p=0}^{n-1}}
l^{2p-2}\frac{\kappa}{2\left(  n-p\right)  -1}\binom{n-1}{p}\mu_{i}%
\delta_{i_{1}+\cdots+i_{2n-1-p}}^{i}\nonumber\\
&  \varepsilon_{a_{1}a_{2}\cdots a_{2n-1}}R^{\left(  a_{1}a_{2},i_{1}\right)
}\cdots R^{\left(  a_{2p-1}a_{2p},i_{p}\right)  }e^{\left(  a_{2p+1}%
,i_{p+1}\right)  }\cdots e^{\left(  a_{2n-1},i_{2n-1-p}\right)  }.
\end{align}
Let us note that this Lagrangian can be expressed equivalently as follows
\footnote{The term with $p=0$ does not contribute to the sum because $\delta
_{i_{1}+\cdots+i_{2n-1}}^{i}=0$ for any value of $i$ and $n$.}
\begin{align}
L_{CS\text{ \ }\left(  2n-1\right)  }^{\mathfrak{B}_{2n-1}}  &  =%
{\displaystyle\sum\limits_{k=1}^{n-1}}
l^{2k-2}c_{k}\alpha_{i}\delta_{i_{1}+\cdots+i_{n}}^{i}\delta_{p_{1}+q_{1}%
}^{i_{k+1}}\cdots\delta_{p_{n-1-k}+q_{n-1-k}}^{i_{n-1}}\nonumber\\
&  \varepsilon_{a_{1}\cdots a_{2n-1}}R^{\left(  a_{1}a_{2},i_{1}\right)
}\cdots R^{\left(  a_{2k-1}a_{2k},i_{k}\right)  }e^{\left(  a_{2k+1}%
,p_{1}\right)  }\nonumber\\
&  e^{\left(  a_{2k+2},q_{1}\right)  }\cdots e^{\left(  a_{2n-3}%
,p_{n-1-k}\right)  }e^{\left(  a_{2n-2},q_{n-1-k}\right)  }e^{\left(
a_{2n-1},i_{n}\right)  },
\end{align}
where
\begin{align}
c_{k}  &  =\frac{1}{2\left(  n-k\right)  -1}\binom{n-1}{k},\\
\alpha_{i}  &  =\kappa\mu_{i}%
\end{align}
and
\begin{equation}
R^{\left(  ab,2i\right)  }=d\omega^{\left(  ab,2i\right)  }+\eta_{cd}%
\omega^{\left(  ac,2j\right)  }\omega^{\left(  db,2k\right)  }\delta_{j+k}%
^{i}.
\end{equation}
This is the Einstein-Chern-Simons Lagrangian $\left[  \text{compare with eq.
}\left(  \ref{ECS}\right)  \right]  $ found in Ref.\cite{salg1}.

\subsection{\textbf{Born-Infeld gravity invariant under }$\mathcal{L}%
^{\mathfrak{B}_{2n}}$}

\qquad In the even-dimensional case, following the same procedure of Ref.
\cite{tron} one can see that eqs. $\left(  \ref{ec3}\right)  $, $\left(
\ref{ec4}\right)  $, lead to the following coefficients%

\begin{equation}
\alpha_{p}=\alpha_{0}\left(  2\gamma\right)  ^{p}\binom{n}{p}.
\end{equation}
With these coefficients the Lagrangian (\ref{lovexp}) is given by%

\begin{align}
L_{BI\text{ \ }\left(  2n\right)  }^{\mathcal{L}^{\mathfrak{B}_{2n}}}  &  =%
{\displaystyle\sum\limits_{p=0}^{n}}
\frac{\kappa}{2n}l^{2p-2}\binom{n}{p}\mu_{i}\delta_{i_{1}+\cdots+i_{2n-p}}%
^{i}\nonumber\\
&  \varepsilon_{a_{1}a_{2}\cdots a_{2n}}R^{\left(  a_{1}a_{2},i_{1}\right)
}\cdots R^{\left(  a_{2p-1}a_{2p},i_{p}\right)  }e^{\left(  a_{2p+1}%
,i_{p+1}\right)  }\cdots e^{\left(  a_{2n},i_{2n-p}\right)  },
\end{align}
or equivalently%
\footnote
{As in the odd-dimensional case $p=0$ does not contribute to the sum because $\delta
_{i_{1}+\cdots+i_{2n}}^{i}=0$ for any value of $i$ and $n$.}%
,%
\begin{align}
L_{BI\text{ \ }(2n)}^{\mathcal{L}^{\mathfrak{B}_{2n}}}  &  =\sum_{k=1}%
^{n}\frac{1}{2n}l^{2k-2}\binom{n}{k}\alpha_{i}\delta_{i_{1}+\cdots+i_{n}}%
^{i}\delta_{p_{1}+q_{1}}^{i_{k+1}}\cdots\delta_{p_{n-k}+q_{n-k}}^{i_{n}%
}\nonumber\\
&  \varepsilon_{a_{1}\cdots a_{2n}}R^{\left(  a_{1}a_{2},i_{1}\right)  }\cdots
R^{\left(  a_{2k-1}a_{2k},i_{k}\right)  }e^{\left(  a_{2k+1},p_{1}\right)
}\nonumber\\
&  e^{\left(  a_{2k+2},q_{1}\right)  }\cdots e^{\left(  a_{2n-1}%
,p_{n-k}\right)  }e^{\left(  a_{2n},q_{n-k}\right)  },
\end{align}
which corresponds to the Einstein-Born-Infeld Lagrangian found in
Ref.\cite{salg1a}. \ It is important to note that the coefficients $\alpha
_{i}=$ $\kappa\mu_{i}$ are arbitrary constants.

In this way we have shown that the $S$-expansion procedure does not modify the
$\alpha_{p}$'s coefficients defined in Ref.\cite{tron}. \ Unlike the
Lanczos-Lovelock action, the expanded action (\ref{lovexp}) called the
Einstein-Lovelock action, has the property of leading to General Relativity in
a certain limit of the coupling constant $l$ both even and odd dimensions.

\section{\textbf{Adding torsion in the Lagrangian }}

\qquad The Lagrangian (\ref{lovexp}) can be interpreted as the most general
$d$-form invariant under a Lorentz type subalgebra $\mathfrak{L}%
^{\mathfrak{B}_{2n}}$ of the generalized Poincar\'{e} algebra. This Lagrangian
is constructed from the expanded vielbein and the expanded spin connection
$e^{\left(  a,2k+1\right)  }$, $\omega^{\left(  ab,2k\right)  }$ $\left(
k=0,\dots,n-1\right)  $ and their exterior derivatives%
\footnote{When $k=0$, $e^{(a,1)}$ and $\omega^{(ab,0)}%
$ are identified with the usual vielbein $e^{a}%
$ and the spin connection $\omega^{ab}$, respectively.}%
.

One can see from the variation of the EL Lagrangian that eq. $\left(
\ref{ec1b}\right)  $ does not imply in $d>4$ the vanishing of the expanded
torsion $T^{\left(  a,2k+1\right)  }$. \ The condition $T^{\left(
a,2k+1\right)  }=0$ implies that the expanded spin connection $\omega^{\left(
ab,2k\right)  }$ have a dependence on the expanded vielbein $e^{\left(
a,2k+1\right)  }$. Thus the expanded fields $\omega^{\left(  ab,2k\right)
}\,$\ and $e^{\left(  a,2k+1\right)  }$ cannot be identified as the components
of a connection for the generalized Poincar\'{e} algebra. \ Therefore, impose
$T^{\left(  a,2k+1\right)  }=0$ seems to be restrictive and arbitrary. \ In
this section, we study the possibility of adding terms which contain\ the
expanded torsion to the $ELC$ Lagrangian.

The Einstein-Lovelock-Cartan Lagrangian can be generalized adding torsion
explicitly following a procedure analogous to that of the Refs. \cite{tron,
Marzan}.

The only terms invariant under $\mathfrak{L}^{\mathfrak{B}_{2n}}$ that can be
constructed out of $e^{\left(  a,2k+1\right)  }$, $\omega^{\left(
ab,2k\right)  }$ and their exterior derivatives, are $R^{\left(  ab,2k\right)
}$, $T^{\left(  a,2k+1\right)  }$, and products of them. \ Then the invariant
combinations that can occur in the Lagrangian are:
\begin{align}
R_{A}^{\left(  i\right)  }  &  =\delta_{2\left(  k_{1}+\cdots+k_{A}\right)
}^{i}R_{\;a_{2}}^{a_{1},\left(  2k_{1}\right)  }\cdots R_{\;a_{1}}%
^{a_{A}\text{ },\left(  2k_{A}\right)  },\\
V_{A}^{\left(  i\right)  }  &  =\delta_{2\left(  k_{1}+\cdots+k_{A}%
+k_{A+1}+k_{A+2}+1\right)  }^{i}R_{\;a_{2}}^{a_{1}\text{ },\left(
2k_{1}\right)  }\cdots R_{\;b}^{a_{A}\text{ },\left(  2k_{A}\right)  }%
e_{a_{1}}^{\text{ },\left(  2k_{A+1}+1\right)  }e^{\left(  b,2k_{A+2}%
+1\right)  },\\
T_{A}^{\left(  i\right)  }  &  =\delta_{2\left(  k_{1}+\cdots+k_{A}%
+k_{A+1}+k_{A+2}+1\right)  }^{i}R_{\;a_{2}}^{a_{1}\text{ },\left(
2k_{1}\right)  }\cdots R_{\;b}^{a_{A}\text{ },\left(  2k_{A}\right)  }%
T_{a_{1}}^{\text{ },\left(  2k_{A+1}+1\right)  }T^{\left(  b,2k_{A+2}%
+1\right)  },\\
K_{A}^{\left(  i\right)  }  &  =\delta_{2\left(  k_{1}+\cdots+k_{A}%
+k_{A+1}+k_{A+2}+1\right)  }^{i}R_{\;a_{2}}^{a_{1}\text{ },\left(
2k_{1}\right)  }\cdots R_{\;b}^{a_{A}\text{ },\left(  2k_{A}\right)  }%
T_{a_{1}}^{\text{ },\left(  2k_{A+1}+1\right)  }e^{\left(  b,2k_{A+2}%
+1\right)  },
\end{align}
where $i=0,\dots,2n-2$. \ So that, the Lagrangian can be written as a linear
combination of products of these basic invariant combinations. In a similar
way to Ref. \cite{Marzan}, we find that the Lagrangian has to be of the form%
\begin{equation}
L=%
{\displaystyle\sum\limits_{p=0}^{\left[  d/2\right]  }}
\mu_{i}\alpha_{p}L_{\mathcal{EL}}^{\left(  p,i\right)  }+%
{\displaystyle\sum\limits_{j}}
\mu_{i}\beta_{j}L_{A_{j}}^{d,\left(  i\right)  },
\end{equation}
where the $\mu,\alpha$ and $\beta$ are constants, $L_{\mathcal{ELC}}^{\left(
p,i\right)  }$ corresponds to the Einstein-Lovelock Lagrangian $\left(
\ref{EL}\right)  $ and $L_{A_{j}}^{d,\left(  i\right)  }$ is a $d$-form
invariant under the $\mathfrak{L}^{\mathfrak{B}}$ algebra given by%
\begin{equation}
L_{A_{j}}^{d,\left(  i\right)  }=R_{A_{1}}^{\left(  i\right)  }\cdots
R_{A_{r}}^{\left(  i\right)  }T_{B_{1}}^{\left(  i\right)  }\cdots T_{B_{t}%
}^{\left(  i\right)  }V_{C_{1}}^{\left(  i\right)  }\cdots V_{C_{v}}^{\left(
i\right)  }K_{D_{1}}^{\left(  i\right)  }\cdots K_{D_{k}}^{\left(  i\right)
}.
\end{equation}

Thus, the inclusion of the expanded torsion leads to a number of arbitrary
coefficients $\beta j$. \ Interestingly, as in the $AdS$ symmetry case, it is
possible to choose the $\beta$'s in order to enlarge the Lorentz type
$\mathfrak{L}^{\mathfrak{B}}$ symmetry to the generalized Poincar\'{e} gauge symmetry.

In even dimensions, the $\mathfrak{B}_{2n+1}$-invariant $d$-forms are given by%
\begin{equation}
\mathcal{P}=\left\langle F^{d/2}\right\rangle ,
\end{equation}
where $\left\langle \dots\right\rangle $ denotes a symmetric invariant tensor
for the $\mathfrak{B}_{2n+1}$ algebra. \ Here, $F=dA+AA$ is the $2$-form
curvature for the generalized Poincar\'{e} algebra and it is given by%
\begin{equation}
F=\sum_{k=0}^{n-1}\left[  \frac{1}{2}F^{\left(  ab,2k\right)  }J_{\left(
ab,2k\right)  }+\frac{1}{l}F^{\left(  a,2k+1\right)  }P_{\left(
a,2k+1\right)  }\right]  ,
\end{equation}
with%
\begin{align}
F^{\left(  ab,2k\right)  }  &  =d\omega^{\left(  ab,2k\right)  }+\eta
_{cd}\omega^{\left(  ac,2i\right)  }\omega^{\left(  db,2j\right)  }%
\delta_{i+j}^{k}+\frac{1}{l^{2}}e^{\left(  a,2i+1\right)  }e^{\left(
b,2j+1\right)  }\delta_{i+j+1}^{k},\\
F^{\left(  a,2k+1\right)  }  &  =de^{\left(  a,2k+1\right)  }+\eta_{bc}%
\omega^{\left(  ab,2i\right)  }e^{\left(  c,2j\right)  }\delta_{i+j}^{k}.
\end{align}
The $\omega^{\left(  ab,2k\right)  }$ and $e^{\left(  a,2k+1\right)  }$ are
the different components of the $1$-form connection $A$,%
\begin{equation}
A=\sum_{k=0}^{n-1}\left[  \frac{1}{2}\omega^{\left(  ab,2k\right)  }J_{\left(
ab,2k\right)  }+\frac{1}{l}e^{\left(  a,2k+1\right)  }P_{\left(
a,2k+1\right)  }\right]  ,
\end{equation}
where $J_{\left(  ab,2k\right)  }$ and $P_{\left(  a,2k+1\right)  }$ are the
generators of the generalized Poincar\'{e} algebra $\mathfrak{B}_{2n+1}$.

Naturally, one of the invariants present in even dimensions is the Euler type
invariant which is obtained from the following components of an invariant
tensor,%
\begin{equation}
\left\langle J_{\left(  a_{1}a_{2},2k_{1}\right)  }\cdots J_{\left(
a_{d-3}a_{d-2},2k_{\left(  d-2\right)  /2}\right)  }P_{\left(  a_{d-1}%
,2k_{d/2}+1\right)  }\right\rangle =\mu_{i}\delta_{2\left(  k_{1}%
+\cdots+k_{d/2}\right)  +1}^{i}\epsilon_{a_{1}a_{2}\cdots a_{d-1}},
\end{equation}
with $k_{i}=0,\cdots,n-1$.

However, there are other components of the invariant tensor which lead to a
different invariant known as the Pontryagin invariant which exists only in
$4p$ dimensions. \ This invariant corresponds to the $\mathfrak{B}_{2n+1}%
$-invariant $d$-form built from $e^{\left(  a,2k+1\right)  },$ $R^{\left(
ab,2k\right)  }$, $T^{\left(  a,2k+1\right)  }$ and can be expressed as the
exterior derivative of a Chern-Simons $\left(  4p-1\right)  $-form,%

\begin{equation}
dL_{T\text{ }\left(  4p-1\right)  }^{\mathfrak{B}_{2n+1}}=P_{\left(
4p\right)  }.
\end{equation}

This implies that in odd dimensions there are two families of Lagrangians
invariant under the generalized Poincar\'{e} algebra $\mathfrak{B}_{2n+1}$: \ 

\begin{itemize}
\item The \textit{Euler-Chern-Simons }form $L_{E\text{ \ }\left(  2p+1\right)
}^{\mathfrak{B}_{2n+1}}$, in $D=2p+1$. \ Its exterior derivative is the Euler
density in $2p+2$ dimensions and does not involve torsion explicitly.

\item The\textit{\ Pontryagin-Chern-Simons }form\textit{ }$L_{T\text{
\ \ }\left(  4p-1\right)  }^{\mathfrak{B}_{2n+1}}$, in $D=4p-1$. \ Its
exterior derivative is the Pontryagin invariant $P_{\left(  4p\right)
}^{\mathfrak{B}_{2n+1}}$in $4p$ dimensions.
\end{itemize}

These results generalize those obtained in Ref. \cite{tron} to our case. \ The
similitude is not a surprise since the $\mathfrak{B}_{2n+1}$ algebra
corresponds to an expansion of the $AdS$ algebra. \ Nevertheless, unlike the
$AdS$-invariant gravity theory, the locally $\mathfrak{B}_{2n+1}$-invariant
gravity theory leads to General Relativity in the weak coupling constant limit
$\left(  l\rightarrow0\right)  $ $\left(  \text{see Ref. \cite{salg1, salg1a,
CPRS2}}\right)  $.

Interestingly, in $4p$ dimensions, both families exist which allows us to
write the most general Lagrangian for gravity in $d=4p-1$ invariant under the
generalized Poincar\'{e} algebra, namely%
\begin{align}
L_{CS\text{ }\left(  4p-1\right)  }^{\mathfrak{B}_{2n+1}}  &  =L_{E\text{
\ }\left(  4p-1\right)  }^{\mathfrak{B}_{2n+1}}+L_{T\text{ \ \ }\left(
4p-1\right)  }^{\mathfrak{B}_{2n+1}}\\
&  =\alpha_{i}L_{E\text{ \ }\left(  4p-1\right)  }^{\left(  i\right)  }%
+\alpha_{j}L_{T\text{ \ \ }\left(  4p-1\right)  }^{\left(  j\right)  },
\end{align}
where $i=1,3,5,\dots,2n-1$ and $j=0,2,4,\dots,2n-2$. \ The $\alpha$'s are
arbitrary and are a consequence of the $S$-expansion procedure. \ In the next
subsection, we explore an example in $d=3$ which clarifies this point.

\subsubsection{\textbf{Example for }$d=3$}

\qquad Let us consider a $(2+1)$-dimensional Lagrangian invariant under the
$\mathfrak{B}_{5}$ algebra. This algebra can be obtained from the $AdS$
algebra, using the $S$-expansion procedure of Ref. \cite{salg2}.

After extracting a resonant subalgebra and performing a $0_{S}$-reduction, one
finds the $\mathfrak{B}_{5}$ algebra, whose generators satisfy the following
commutation relations%
\begin{align}
\left[  P_{a},P_{b}\right]   &  =Z_{ab},\text{ \ \ \ \ }\left[  J_{ab}%
,P_{c}\right]  =\eta_{bc}P_{a}-\eta_{ac}P_{b}\\
\left[  J_{ab,}J_{cd}\right]   &  =\eta_{cb}J_{ad}-\eta_{ca}J_{bd}+\eta
_{db}J_{ca}-\eta_{da}J_{cb}\\
\left[  J_{ab,}Z_{cd}\right]   &  =\eta_{cb}Z_{ad}-\eta_{ca}Z_{bd}+\eta
_{db}Z_{ca}-\eta_{da}Z_{cb}\\
\left[  J_{ab},Z_{c}\right]   &  =\eta_{bc}Z_{a}-\eta_{ac}Z_{b},\text{
\ \ \ \ }\\
\left[  Z_{ab},P_{c}\right]   &  =\eta_{bc}Z_{a}-\eta_{ac}Z_{b},\text{
\ \ \ \ }\\
\left[  Z_{ab},Z_{c}\right]   &  =\left[  Z_{ab,}Z_{cd}\right]  =\left[
P_{a},Z_{c}\right]  =0.
\end{align}

In order to write down a Chern-Simons Lagrangian for the $\mathfrak{B}_{5}$
algebra, we start from the $\mathfrak{B}_{5}$-valued one-form gauge connection%
\begin{equation}
A=\frac{1}{2}\omega^{ab}J_{ab}+\frac{1}{l}e^{a}P_{a}+\frac{1}{2}k^{ab}%
Z_{ab}+\frac{1}{l}h^{a}Z_{a},
\end{equation}
and the associated two-form curvature%
\begin{equation}
F=\frac{1}{2}R^{ab}J_{ab}+\frac{1}{l}T^{a}P_{a}+\frac{1}{2}\left(  D_{\omega
}k^{ab}+\frac{1}{l^{2}}e^{a}e^{b}\right)  Z_{ab}+\frac{1}{l}\left(  D_{\omega
}h^{a}+k_{\text{ }b}^{a}e^{b}\right)  Z_{a}.
\end{equation}

Using Theorem VII.2 of Ref. \cite{salg2}, it is possible to show that the only
non-vanishing components of an invariant tensor for the $\mathfrak{B}_{5}$
algebra are given by%
\begin{align}
\left\langle J_{ab}J_{cd}\right\rangle _{\mathfrak{B}_{5}}  &  =\alpha
_{0}\left(  \eta_{ad}\eta_{bc}-\eta_{ac}\eta_{bd}\right)  ,\text{
\ \ \ \ \ \ }\left\langle J_{ab}P_{c}\right\rangle _{\mathfrak{B}_{5}}%
=\alpha_{1}\epsilon_{abc},\\
\left\langle J_{ab}Z_{cd}\right\rangle _{\mathfrak{B}_{5}}  &  =\alpha
_{2}\left(  \eta_{ad}\eta_{bc}-\eta_{ac}\eta_{bd}\right)  ,\text{
\ \ \ \ \ \ }\left\langle J_{ab}Z_{c}\right\rangle _{\mathfrak{B}_{5}}%
=\alpha_{3}\epsilon_{abc},\\
\left\langle P_{a}P_{c}\right\rangle _{\mathfrak{B}_{5}}  &  =\alpha_{2}%
\eta_{ac},\text{ \ \ \ \ \ \ \ \ \ \ \ \ \ \ \ \ \ \ \ \ \ \ \ \ \ }%
\left\langle Z_{ab}P_{c}\right\rangle _{\mathfrak{B}_{5}}=\alpha_{3}%
\epsilon_{abc},
\end{align}
where $\alpha_{0},$ $\alpha_{1},$ $\alpha_{2}$ and $\alpha_{3}$ are arbitrary constants.

Using these components of the invariant tensor in the general expression for
the $ChS$ Lagrangian $L_{ChS}=\left\langle AdA+\frac{2}{3}A^{3}\right\rangle
$, we find that the $ChS$ Lagrangian invariant under the $\mathfrak{B}_{5}$
algebra is given by%
\begin{align}
L_{ChS\text{ \ }\left(  2+1\right)  }^{\mathfrak{B}_{5}}  &  =\frac{1}%
{l}\varepsilon_{abc}\left[  \alpha_{1}R^{ab}e^{c}+\alpha_{3}\left(  \frac
{1}{3l^{2}}e^{a}e^{b}e^{c}+R^{ab}h^{c}+k^{ab}T^{c}\right)  \right] \nonumber\\
&  +\frac{\alpha_{0}}{2}\left(  \omega_{\;b}^{a}d\omega_{\;a}^{b}+\frac{2}%
{3}\omega_{\hspace{0.05cm}b}^{a}\omega_{\hspace{0.05cm}c}^{b}\omega
_{\hspace{0.05cm}a}^{c}\right)  +\frac{\alpha_{2}}{2}\left(  \frac{2}{l^{2}%
}e^{a}T_{a}+\omega_{\;b}^{a}dk_{\;a}^{b}+k_{\;b}^{a}d\omega_{\;a}^{b}%
+2\omega_{\hspace{0.05cm}b}^{a}\omega_{\hspace{0.05cm}c}^{b}k_{\hspace
{0.05cm}a}^{c}\right) \label{CSLB5}\\
&  =\alpha_{1}L_{E\text{ \ }\left(  3\right)  }^{\left(  1\right)  }%
+\alpha_{3}L_{E\text{ \ }\left(  3\right)  }^{\left(  3\right)  }+\alpha
_{0}L_{T\text{ \ \ }\left(  3\right)  }^{\left(  0\right)  }+\alpha
_{2}L_{T\text{ \ \ }\left(  3\right)  }^{\left(  2\right)  }.
\end{align}

The exterior derivative of this Lagrangian leads us to the following
associated invariant%
\begin{align}
\mathcal{P}_{\text{ }\left(  4\right)  }^{\mathfrak{B}_{5}}  &  =\frac{1}%
{l}\varepsilon_{abc}\left[  \alpha_{1}R^{ab}T^{c}+\alpha_{3}\left(  \frac
{1}{l^{2}}e^{a}e^{b}T^{c}+R^{ab}\left(  D_{\omega}h^{c}+k_{\text{ \ }d}%
^{c}e^{d}\right)  +D_{\omega}k^{ab}T^{c}\right)  \right] \nonumber\\
&  +\frac{\alpha_{0}}{2}R_{\text{ \ }b}^{a\text{ }}R_{\text{ \ }a}^{b}%
+\frac{\alpha_{2}}{2}\left[  \frac{2}{l^{2}}\left(  T^{a}T_{a}-e^{a}%
e^{b}R_{ab}\right)  +2R_{\text{ \ }b}^{a\text{ \ }}D_{\omega}k_{\text{ \ }%
a}^{b}\right]  .
\end{align}
where in addition to an Euler type density we can see that appears the usual
Pontryagin density $P_{\left(  4\right)  }=R_{\text{ \ }b}^{a\text{ }%
}R_{\text{ \ }a}^{b}$, the Nieh-Yan $N_{\left(  4\right)  }=$ $\frac{2}{l^{2}%
}\left(  T^{a}T_{a}-e^{a}e^{b}R_{ab}\right)  $ and a Pontryagin type density
$P_{4}\left(  k\right)  =2R_{\text{ \ }b}^{a\text{ \ }}D_{\omega}k_{\text{
\ }a}^{b}$ coming from the new fields.

Note that these densities $P_{\left(  4\right)  }$, $N_{\left(  4\right)  }$
and $P_{\left(  4\right)  }\left(  k\right)  $ are combined in a Pontryagin
type invariant for the $B_{5}$ group which is written as follows (choosing
$\alpha_{0}=\alpha_{2}$)%
\begin{equation}
F_{\text{ }B}^{A}F_{\text{ }A}^{B}=R_{\hspace{0.05cm}b}^{a}R_{\hspace
{0.05cm}a}^{b}+\left[  \frac{2}{l^{2}}\left(  T^{a}T_{a}-e^{a}e^{b}%
R_{ab}\right)  +2R_{\text{ \ }b}^{a\text{ \ }}D_{\omega}k_{\text{ \ }a}%
^{b}\right]  ,
\end{equation}
where%
\begin{equation}
F^{AB}=\left(
\begin{array}
[c]{cc}%
\begin{array}
[c]{c}%
R^{ab}+\left(  D_{\omega}k^{ab}+\frac{1}{l^{2}}e^{a}e^{b}\right) \\
\end{array}
&
\begin{array}
[c]{c}%
\frac{1}{l}T^{a}+\frac{1}{l}\left(  D_{\omega}h^{a}+k_{\text{ }c}^{a}%
e^{c}\right) \\
\end{array}
\\%
\begin{array}
[c]{c}%
-\frac{1}{l}T^{b}-\frac{1}{l}\left(  D_{\omega}h^{b}+k_{\text{ }c}^{b}%
e^{c}\right) \\
\end{array}
&
\begin{array}
[c]{c}%
0\\
\end{array}
\end{array}
\right)
\end{equation}

In the next section we show that the $\mathfrak{B}_{5}$-invariant Lagrangian
$\left(  \ref{CSLB5}\right)  $ can be obtained directly from the
Lorentz-invariant Lagrangian.

\section{\textbf{Relation between the Pontryagin and Euler invariants}}

In this section we show that it is possible to relate the Lorentz invariant
Lagrangian which depends only on the spin connection, and the Lagrangian
obtained for the $\mathfrak{B}_{5}$ algebra . This means that by dual
formulation of the $S$-expansion \cite{salg3} is possible to obtain both an
Euler type invariant and a Pontryagin type invariant from the Pontryagin invariant.

Consider first the Lorentz algebra $\mathcal{L}$ in $\left(  2+1\right)  $-dimensions,%

\begin{equation}
\left[  J_{ab},J_{cd}\right]  =\eta_{cb}J_{ad}-\eta_{ca}J_{bd}+\eta_{db}%
J_{ca}-\eta_{da}J_{cb}%
\end{equation}

The one-form gauge connection $A$ and the associated two-form curvature $F$
are given by%

\begin{align}
A  &  =\frac{1}{2}\omega^{ab}J_{ab},\\
F  &  =\frac{1}{2}R^{ab}J_{ab},
\end{align}
where $R^{ab}=d\omega^{ab}+\omega_{\text{ }c}^{a}\omega^{cb}$ is the Lorentz
curvature. The corresponding Chern-Simons Lagrangian invariant under the
Lorentz algebra $\mathcal{L}$ is given by%
\begin{equation}
L_{3}^{Lorentz}={\normalsize \omega}_{\hspace{0.05cm}b}^{a}d\omega
_{\hspace{0.05cm}a}^{b}+\frac{2}{3}\omega_{\hspace{0.05cm}b}^{a}%
\omega_{\hspace{0.05cm}c}^{b}\omega_{\hspace{0.05cm}a}^{c},
\end{equation}
which can be written as%
\[
L_{3}^{Lorentz}=\left\langle AdA+\frac{2}{3}A^{3}\right\rangle ,
\]
where the invariant tensor $\left\langle \cdots\right\rangle $ for the Lorentz
algebra is%
\[
\left\langle J_{ab}J_{cd}\right\rangle _{\mathcal{L}}=\eta_{ad}\eta_{bc}%
-\eta_{ac}\eta_{bd}.
\]

Before starting the $S$-expansion of the Lorentz algebra is useful to define%

\[
J^{a}=-\frac{1}{2}\epsilon^{abc}J_{bc},\text{ \ \ \ \ }\omega_{a}=-\frac{1}%
{2}\epsilon_{abc}\omega^{bc},
\]
so that%
\begin{equation}
A=\omega_{a}J^{a},\text{ \ \ \ \ \ }F=F_{a}J^{a},
\end{equation}
with%
\begin{equation}
F_{a}=-\frac{1}{2}\epsilon_{abc}R^{bc}=d\omega_{a}-\frac{1}{2}\eta
_{ab}\epsilon^{bcd}\omega_{c}\omega_{d}.
\end{equation}

Now let us consider the $S_{E}^{\left(  3\right)  }$ expansion of Lorentz
algebra. The appropriate semigroup $S_{E}^{\left(  3\right)  }=$ $\left\{
\lambda_{0},\lambda_{1},\lambda_{2},\lambda_{3},\lambda_{4}\right\}  $ is
endowed with the following product:%

\begin{equation}
\lambda_{\alpha}\lambda_{\beta}=\left\{
\begin{array}
[c]{c}%
\lambda_{\alpha+\beta},\text{ \ when }\alpha+\beta\leq4\\
\lambda_{4},\text{ \ \ \ \ when }\alpha+\beta>4
\end{array}
\right.
\end{equation}
where $\lambda_{4}=0_{s}$ is the zero element of the semigroup.

In a similar way to Ref. \cite{salg3}, we define the spin connection and the
2-form curvature as
\begin{align}
\omega_{a}  &  =\lambda_{0}\omega_{a}^{\text{ }\left(  0\right)  }+\lambda
_{1}\omega_{a}^{\text{ }\left(  1\right)  }+\lambda_{2}\omega_{a}^{\text{
}\left(  2\right)  }+\lambda_{3}\omega_{a}^{\text{ }\left(  3\right)  },\\
F_{a}  &  =\lambda_{0}F_{a}^{\text{ }\left(  0\right)  }+\lambda_{1}%
F_{a}^{\text{ }\left(  1\right)  }+\lambda_{2}F_{a}^{\text{ }\left(  2\right)
}+\lambda_{3}F_{a}^{\text{ }\left(  3\right)  },
\end{align}
where%
\begin{align*}
\omega_{a}^{\text{ }\left(  0\right)  }  &  =\omega_{a},\text{ \ }\omega
_{a}^{\text{ }\left(  2\right)  }=k_{a},\\
\omega_{a}^{\text{ }\left(  1\right)  }  &  =\frac{1}{l}e_{a},\text{ \ }%
\omega_{a}^{\text{ }\left(  3\right)  }=\frac{1}{l}h_{a},
\end{align*}
and%
\begin{align}
F_{a}^{\text{ }\left(  0\right)  }  &  =-\frac{1}{2}\epsilon_{abc}R^{bc},\\
F_{a}^{\text{ }\left(  1\right)  }  &  =\frac{1}{l}T_{a},\\
F_{a}^{\text{ }\left(  2\right)  }  &  =-\frac{1}{2}\epsilon_{abc}\left(
D_{\omega}k^{bc}+\frac{1}{l^{2}}e^{b}e^{c}\right)  ,\\
F_{a}^{\text{ }\left(  3\right)  }  &  =\frac{1}{l}\left(  D_{\omega}%
h^{a}+k_{\text{ \ }b}^{a}e^{b}\right)  .
\end{align}
Here we identify $e^{a}$ with the vielbein, $R^{ab}$ with the Lorentz
curvature , $T^{a}$ with the torsion, and $k^{ab}$ and $h^{a}$ are identified
as bosonic \textquotedblleft matter\textquotedblright\ fields.

\ Using Theorem VII.2 of Ref. \cite{salg2}, it is possible to show that the
only non-vanishing components of an invariant tensor for the $\mathfrak{B}%
_{5}$ algebra are given by%
\begin{align}
\left\langle J_{ab}J_{cd}\right\rangle \text{\ }  &  =\alpha_{0}\left(
\eta_{ad}\eta_{bc}-\eta_{ac}\eta_{bd}\right)  ,\text{
\ \ \ \ \ \ \ \ \ \ \ \ \ \ \ \ \ \ }\label{T1}\\
\left\langle J_{ab}Z_{cd}\right\rangle  &  =\alpha_{2}\left(  \eta_{ad}%
\eta_{bc}-\eta_{ac}\eta_{bd}\right)  ,\\
\left\langle P_{a}P_{c}\right\rangle  &  =\alpha_{2}\eta_{ac},\\
\left\langle J_{ab}P_{c}\right\rangle  &  =\alpha_{1}\epsilon_{abc},\\
\left\langle J_{ab}Z_{c}\right\rangle  &  =\alpha_{3}\epsilon_{abc},\\
\left\langle Z_{ab}P_{c}\right\rangle  &  =\alpha_{3}\epsilon_{abc},
\label{T6}%
\end{align}
where $\alpha_{0}$, $\alpha_{1}$, $\alpha_{2}$ and $\alpha_{3}$ are arbitrary
constants. \ Now if we use the components of the invariant tensor $\left(
\ref{T1}\right)  -\left(  \ref{T6}\right)  $ in the general expression for a
Chern-Simons Lagrangian we find the $\mathfrak{B}_{5}$-invariant CS Lagrangian
in $\left(  2+1\right)  $ dimensions,%

\begin{align}
L_{CS\text{ \ }\left(  2+1\right)  }^{\mathfrak{B}_{5}}  &  =\frac{1}%
{l}\varepsilon_{abc}\left[  \alpha_{1}R^{ab}e^{c}+\alpha_{3}\left(  \frac
{1}{3l^{2}}e^{a}e^{b}e^{c}+R^{ab}h^{c}+k^{ab}T^{c}\right)  \right] \nonumber\\
&  +\frac{\alpha_{0}}{2}\left(  \omega_{\;b}^{a}d\omega_{\;a}^{b}+\frac{2}%
{3}\omega_{\hspace{0.05cm}b}^{a}\omega_{\hspace{0.05cm}c}^{b}\omega
_{\hspace{0.05cm}a}^{c}\right)  +\frac{\alpha_{2}}{2}\left(  \frac{2}{l^{2}%
}e^{a}T_{a}+\omega_{\;b}^{a}dk_{\;a}^{b}+k_{\;b}^{a}d\omega_{\;a}^{b}%
+2\omega_{\hspace{0.05cm}b}^{a}\omega_{\hspace{0.05cm}c}^{b}k_{\hspace
{0.05cm}a}^{c}\right)  . \label{lag3a}%
\end{align}
The exterior derivative of this Lagrangian leads us to the following invariant polynomial,%

\begin{align}
P_{\text{ }\left(  4\right)  }^{\mathfrak{B}_{5}}  &  =\frac{1}{l}%
\epsilon_{abc}\left[  \alpha_{1}R^{ab}T^{c}+\alpha_{3}\left(  \frac{1}{l^{2}%
}e^{a}e^{b}T^{c}+R^{ab}\left(  D_{\omega}h^{c}+k_{\text{ \ }d}^{c}%
e^{d}\right)  +D_{\omega}k^{ab}T^{c}\right)  \right] \nonumber\\
&  +\frac{\alpha_{0}}{2}R_{\text{ \ }b}^{a\text{ }}R_{\text{ \ }a}^{b}%
+\frac{\alpha_{2}}{2}\left[  \frac{2}{l^{2}}\left(  T^{a}T_{a}-e^{a}%
e^{b}R_{ab}\right)  +2R_{\text{ \ }b}^{a\text{ \ }}D_{\omega}k_{\text{ \ }%
a}^{b}\right]  . \label{lag3b}%
\end{align}

Thus we have shown that the $S$-expansion method allows us to relate the
Pontryagin invariant of the Lorentz algebra with the invariants of the
$\mathfrak{B}_{5}$ algebra studied in the previous section.

It is important to note that it is possible to generalize the previous result
to the case of the $\mathfrak{B}_{2n+1}$ algebras. \ In fact, by considering
the reduced $S_{E}^{\left(  2n-1\right)  }$-expansion of the Lorentz algebra
$\mathcal{L}$ and using the Theorem VII.2 of Ref. \cite{salg2} we can find the
non-vanishing components of an invariant tensor for the expanded algebra and
thus build a $\left(  2+1\right)  $-dimensional Lagrangian invariant under
$\mathfrak{B}_{2n+1}$.

\section{\textbf{Comment and possible developments}}

\qquad In the present work we have shown that it is possible to construct an
Einstein-Lovelock-Cartan Lagrangian that, in odd dimensions leads to the
Einstein-Chern-Simons Lagrangian, and in even dimensions leads to the
Einstein-Born-Infeld Lagrangian. \ The $EChS$ and $EBI$ theories are
particularly interesting since it was shown in Refs. \cite{salg1, salg1a,
CPRS2} that General Relativity can be obtained as a certain limit of these
gravity theories.\ On the other hand we have shown that the
Einstein-Lovelock-Cartan Lagrangian can be generalized adding torsional terms
following a procedure analogous to that of Ref. \ \cite{tron}.
\ Interestingly, the torsional terms appear explicitly in the Lagrangian only
in $4p-1$ dimensions. \ Thus, the only $4p$-forms invariant under the
generalized Poincar\'{e} algebra $\mathfrak{B}_{2n+1}$, constructed from
$e^{\left(  a,2k+1\right)  }$, $R^{\left(  ab,2k\right)  }$ and $T^{\left(
a,2k+1\right)  }$ $\left(  k=0,\cdots,n-1\right)  $, are the Pontryagin
invariants $P_{\left(  4p\right)  }$. \ Finally we have established a relation
between the Pontryagin and the Euler invariants using the dual formulation of
the $S$-expansion method introduced in Ref. \cite{salg3}.

The procedure considered here could play an important role in the context of
supergravity in higher dimensions. In fact, it seems likely that it is
possible to recover the standard odd and even-dimensional\ supergravity from a
Chern-Simons and Born-Infeld gravity theories, in a way very similar to the
one shown here. In this way, the procedure sketched here could provide us with
valuable information of what the underlyng geometric structure of Supergravity
could be (work in progress).

\begin{acknowledgement}
This work was supported in part by FONDECYT Grants N$^{0}$ 1130653. Three of
the authors (PKC, DM, EKR) were supported by grants from the Comisi\'{o}n
Nacional de Investigaci\'{o}n Cient\'{\i}fica y Tecnol\'{o}gica CONICYT and
from the Universidad de Concepci\'{o}n, Chile.
\end{acknowledgement}

\end{document}